\documentclass[square]{ws-procs975x65}

\def\idty{{\mathchoice {\mathrm{1\mskip-4mu l}} {\mathrm{1\mskip-4mu l}} 
{\mathrm{1\mskip-4.5mu l}} {\mathrm{1\mskip-5mu l}}}}

\begin{document}

\title{Lieb-Robinson Bounds and Quasi-locality \\
for the Dynamics of Many-Body Quantum Systems}

\author{Robert Sims}

\address{Department of Mathematics, University of Arizona,\\
Tucson, AZ 85721, USA\\
$^*$E-mail: rsims@math.arizona.edu\\
http://math.arizona.edu/~rsims/}

\begin{abstract}
We review a recently proven Lieb-Robinson bound for general, many-body
quantum systems with bounded interactions. Several basic examples are
discussed as well as the connection between commutator estimates and
quasi-locality.
\end{abstract}


\bodymatter

\section{Introduction} 
Much physical intuition is based on locality properties of the
system under consideration. Objects (or particles) are associated
with regions (or points) in space and non-trivial interactions
typically occur over short distances. For systems governed by
a relativistic dynamics, the time evolution preserves this notion of 
locality. 

In condensed matter physics, however, many interesting physically
relevant phenomena are modeled by quantum many-body systems, e.g., 
super-conductivity and Bose-Einstein condensation. More abstractly, 
the models of quantum computation and discrete versions of
field theory are described in terms of quantum lattice systems. 
For these non-relativistic systems, defined e.g. by a Hamiltonian
with nearest neighbor interactions, the associated dynamics 
does not preserve locality in the sense that there is no
strict equivalent to a finite speed of light. 

Remarkably, Lieb and Robinson, see \cite{lieb1972}, proved that an approximate
form of locality, which we refer to as {\it quasi-locality}, does hold
for the dynamics associated to certain quantum spin systems. 
This important result establishes the existence and proves a bound
for an approximate light cone which limits the rate at which disturbances, as
evolved by the dynamics, can propagate through the system.
More concretely, they proved that a local observable evolved 
for a time $t>0$ remains essentially localized to a region of
space whose diameter is proportional to $t$.
They dubbed their estimate as a bound on the system's
group velocity, but we prefer to describe the analysis as a Lieb-Robinson bound
and the resulting estimate: the Lieb-Robinson velocity.

After the initial result by Lieb and Robinson in 1972 and some calculations for
specific models \cite{radin1978} a few years later, these locality estimates for quantum 
systems received relatively little attention. It was not until Hastings' impressive 
work of 2004, see \cite{hastings2004a}, that a genuinely renewed interest in these bounds was
established. Since then, a number of generalizations of the original result 
\cite{NachS06,  hastings06, Nach06, cramer:2008, NachS09, Nach09, PS09, Poulin10}
and a wealth of interesting applications 
\cite{hastings2004a, hastings05, bravyi:2006, hastings2007, Nach07, Burrell07, Osborne07, Erdos08, Hamza09, hastings09, Amour09, debeaudrap10, bravyi:2010a, bravyi:2010b, hastings2010, matsui:2010, Nach10} have demonstrated the importance of these bounds.   
  
In this brief note, we introduce the general set-up for Lieb-Robinson
bounds in the context of quantum systems with bounded interactions.
We describe the correspondence between quasi-locality and the 
usual commutator estimates typically referred to as Lieb-Robinson bounds. 
The final section is a short list of further generalizations and a variety of applications.

Before we begin, we make the following useful observation. In quantum mechanics,
one is often interested in a single quantum system, i.e., a specific
Hilbert space $\mathcal{H}$ and a densely defined, self-adjoint operator
$H$. For each normalized vector, or state $\psi \in \mathcal{H}$, the
solution of the Schr\"odinger equation
\begin{equation}
\partial_t \psi = - i H \psi
\end{equation}
governs the dynamics of this system. The solution,  of course, is $\psi(t) = e^{-itH} \psi$.
For particles in a domain $\Lambda \subset \mathbb{R}^{\nu}$, the typical
Hilbert space is $\mathcal{H} = L^2( \Lambda)$ and it is common to have $H$ a self-adjoint
realization of the Laplacian. Corresponding to any normalized vector $\psi \in L^2( \Lambda)$
the solution $\psi(t)$ is called a wave function, and it is interesting to
investigate the evolution of this function in space, i.e. $\Lambda$.

This is {\it not} the form of locality established by a Lieb-Robinson bound.
Lieb-Robinson bounds  are about collections of interacting quantum systems
distributed in space. The bounds estimate the rate at which disturbances propagate
through this collection.

\section{The General Set-Up}
As discussed above, Lieb-Robinson bounds estimate the rate at which
disturbances propagate through a collection of quantum systems. The basic set-up
is as follows. 

\subsection{Collections of Quantum Systems} 
Let $\Gamma$ be a countable set, and consider a collection of quantum 
systems labeled by $x \in \Gamma$. By this, we mean that corresponding
to each site $x \in \Gamma$ there is a Hilbert space $\mathcal{H}_x$  and
a densely defined, self-adjoint operator $H_x$ acting on $\mathcal{H}_x$.
The operator $H_x$ is typically referred to as the on-site Hamiltonian. 
For finite $\Lambda \subset \Gamma$, the Hilbert space of
states corresponding to $\Lambda$ is given by
\begin{equation}
\mathcal{H}_{\Lambda} = \bigotimes_{x \in \Lambda} \mathcal{H}_x \, ,
\end{equation}
and the algebra of observables is
\begin{equation}
\mathcal{A}_{\Lambda} = \bigotimes_{x \in \Lambda} \mathcal{B}( \mathcal{H}_x) = \mathcal{B}( \mathcal{H}_{\Lambda}) \, ,
\end{equation}
where $\mathcal{B}( \mathcal{H})$ is the set of bounded linear operators over the Hilbert
space $\mathcal{H}$. Thus an observable $A \in \mathcal{A}_{\Lambda}$
depends only on those degrees of freedom in $\Lambda$. Of course, for
any finite $X \subset \Lambda$, an observable $A \in \mathcal{A}_X$ 
can be uniquely identified with the observable 
$A \otimes \idty_{\Lambda \setminus X} \in \mathcal{A}_{\Lambda}$,
and therefore $\mathcal{A}_X \subset \mathcal{A}_{\Lambda}$.  

In general, these collections of quantum systems are used to describe many
interesting physical phenomena e.g., the moments associated with atoms in a 
magnetic material, a lattice of coupled oscillators, or an array of qubits in which
quantum information is stored. Below we indicate two important types of examples.

\begin{example} A {\it quantum spin system} over $\Gamma$ is defined by 
associating a finite dimensional Hilbert space to each site 
$x \in \Gamma$, e.g., $\mathcal{H}_x = \mathbb{C}^{n_x}$ for some integer $n_x \geq 2$. 
The dimension of $\mathcal{H}_x$ is related to the spin at site $x$ by $n_x = 2J_x+1$, i.e.
$n_x=2$ corresponds to spin $J_x=1/2$, $n_x=3$ corresponds to spin $J_x=1$, etc. 
As an on-site Hamiltonian, a common choice is to select a spin matrix in the
$n_x$-dimensional irreducible representation of $su(2)$. When $n_x=2$, these
are just the Pauli spin matrices:
\begin{equation} \label{eq:pauli}
S^1 = \left( \begin{array}{cc} 0 & 1 \\ 1 & 0 \end{array} \right), \quad  S^2 = \left( \begin{array}{cc} 0 & -i \\ i & 0 \end{array} \right), \quad \mbox{and} \quad
S^3 = \left( \begin{array}{cc} 1 & 0 \\ 0 & -1 \end{array} \right).
\end{equation}
\end{example}

\begin{example} A {\it quantum oscillator system} over $\Gamma$ corresponds to 
associating an $L^2$ space to each site of $\Gamma$, e.g. one can take 
$\mathcal{H}_x = L^2( \mathbb{R})$ for each $x \in \Gamma$. In contrast to
the previous example, each single site Hilbert space is infinite dimensional, and moreover,
the on-site Hamiltonians are typically functions of position $q_x$, the multiplication operator
by $q_x$ in $L^2( \mathbb{R}, dq_x)$,  and momentum $p_x = -i \frac{d}{d q_x}$; both unbounded self-adjoint operators.  
\end{example}

Despite the fact that these examples are quite different,  
the general techniques described below apply equally well to both cases.  

\subsection{Interactions and Models} \label{sec:IaM}
The systems described above are of particular interest when they
are allowed to interact. In general, a {\it bounded interaction} for
such quantum systems is a mapping $\Phi$ from the set of finite
subsets of $\Gamma$ into the algebra of observables which satisfies
\begin{equation}
\Phi(X)^* = \Phi(X) \in \mathcal{A}_X \quad \mbox{for all finite } X \subset \Gamma. 
\end{equation} 
A {\it model} is defined by the set $\Gamma$, the collection of 
quantum systems $\{ ( \mathcal{H}_x, H_x ) \}_{x \in \Gamma}$, and an
interaction $\Phi$. 

Associated to a given model there is a family of {\it local Hamiltonians}, $\{ H_{\Lambda} \}$,
parametrized by the finite subsets of $\Gamma$. In fact, to 
each finite $\Lambda \subset \Gamma$, 
\begin{equation}
H_{\Lambda} = \sum_{x \in \Lambda} H_x + \sum_{X \subset \Lambda} \Phi(X)
\end{equation} 
is a densely defined, self-adjoint operator. Here the second sum is over all
finite subsets of $\Lambda$, and is therefore finite. By Stone's theorem, the
corresponding {\it Heisenberg dynamics}, $\tau_t^{\Lambda}$, given by
\begin{equation}
\tau_t^{\Lambda}(A) = e^{itH_{\Lambda}} A e^{-it H_{\Lambda}} \quad \mbox{ for all } A \in \mathcal{A}_{\Lambda} \mbox{ and } t \in \mathbb{R}, 
\end{equation}
is a well-defined, one-parameter group of automorphisms on $\mathcal{A}_{\Lambda}$. 
Here are two typical models of interest.

\begin{example} \label{ex:HH} Fix an integer $\nu \geq 1$ and let $\Gamma = \mathbb{Z}^{\nu}$.
Consider the quantum spin system obtained by setting $\mathcal{H}_x = \mathbb{C}^2$ for
all $x \in \mathbb{Z}^{\nu}$. Take as on-site Hamiltonian $H_x = S^3$ using the notation from 
(\ref{eq:pauli}) above. Let $\Phi$ be the interaction defined by setting
\begin{equation}
\Phi(X) = \left\{ \begin{array}{cc}  S^1_x S^1_y + S^2_x S^2_y + S^3_x S^3_y  & \quad \quad \mbox{if } X = \{ x, y \} \mbox{ and } |x-y|=1, \\
0 & \mbox{otherwise} \end{array} \right.
\end{equation}
where for each $z \in \mathbb{Z}^{\nu}$, any $k \in \{ 1,2,3\}$, and each finite volume $\Lambda \subset \mathbb{Z}^{\nu}$, 
the quantity
\begin{equation}
S_z^k = \idty \otimes \cdots \otimes \idty \otimes S^k \otimes \idty \otimes \cdots \otimes \idty  
\end{equation}
where $S^k$, again from (\ref{eq:pauli}), appears in the $z$-th factor of 
$\mathcal{A}_{\Lambda} = \bigotimes_{x \in \Lambda} \mathcal{B}( \mathbb{C}^2)$.

A nearest neighbor, spin 1/2 Heisenberg model on $\mathbb{Z}^{\nu}$ corresponds to 
\begin{eqnarray}
H_{\Lambda} & = & h \sum_{x \in \Lambda} H_x + J \sum_{X \subset \Lambda} \Phi(X) \nonumber \\
& = & \sum_{x \in \Lambda} h S_x^3 + \sum_{\stackrel{x,y \in \Lambda:}{|x-y|=1}} J \left( S^1_x S^1_y + S^2_x S^2_y + S^3_x S^3_y \right)
\end{eqnarray}
for all finite subsets $\Lambda \subset \mathbb{Z}^{\nu}$. Here $h$ and $J$ are real-valued parameters of the
model. 
\end{example}

\begin{example} \label{ex:AH} Fix an integer $\nu \geq 1$ and let $\Gamma = \mathbb{Z}^{\nu}$.
Consider the quantum oscillator system obtained by setting $\mathcal{H}_x = L^2(\mathbb{R})$ for
all $x \in \mathbb{Z}^{\nu}$.  A nearest neighbor, anharmonic model on 
$\mathbb{Z}^{\nu}$ is defined analogously, e.g. with 
\begin{equation}
H_{\Lambda} = \sum_{x \in \Lambda} p_x^2 +V(q_x)  + \sum_{\stackrel{x,y \in \Lambda:}{|x-y|=1}} \Phi(q_x-q_y)
\end{equation}
for all finite subsets $\Lambda \subset \mathbb{Z}^{\nu}$. The parameters of this model are
$V$ and $\Phi$. Of course, $V$ must be chosen so that the on-site Hamiltonian
$H_x = p_x^2 + V(q_x)$ is self-adjoint, and $\Phi$ is assumed to be in $L^{\infty}( \mathbb{R})$.
\end{example}

\subsection{Observables and Support}

The support of an observable plays a crucial role in Lieb-Robinson bounds. 
We introduce this notion here. Let $\Gamma$ be a countable set and 
$\{ (\mathcal{H}_x, H_x) \}_{x \in \Gamma}$ a collection of quantum systems. 
As we have seen above, for any two finite sets $\Lambda_0 \subset \Lambda \subset \Gamma$, each
$A \in \mathcal{A}_{\Lambda_0}$ can be identified with a unique
element $A \otimes \idty_{\Lambda \setminus \Lambda_0} \in \mathcal{A}_{\Lambda}$.
For this reason, $\mathcal{A}_{\Lambda_0} \subset \mathcal{A}_{\Lambda}$ for any $\Lambda_0 \subset \Lambda$. 

Given an observable $A \in \mathcal{A}_{\Lambda}$, we say that $A$ is {\it supported} in $X \subset \Lambda$ if
$A$ can be written as $A = \tilde{A} \otimes \idty_{\Lambda \setminus X}$ with $\tilde{A} \in \mathcal{A}_X$. The {\it support}
of an observable $A$ is then the minimal set $X$ such that $A$ is supported in $X$. We will denote the support of
an observable $A$ by $\mbox{supp}(A)$. 

Due to the fact that we are considering non-relativistic systems, i.e., models
for which there is no strict equivalent to a finite speed of light, the following
observation is generally true. Let $X \subset \Lambda \subset \Gamma$.
Consider a local Hamiltonian $H_{\Lambda}$ defined in terms of
a non-trivial interaction; assume e.g. the interaction is nearest neighbor. 
Then, for general $A \in \mathcal{A}_X$,
$\mbox{supp}( \tau_0(A) ) = \mbox{supp}(A) \subset X$, however,  $\mbox{supp}( \tau_t^{\Lambda}(A) ) = \Lambda$ for all $ t \neq 0$.
Hence, a strict notion of {\it locality}, implicitly defined here in terms of the support of an observable, is
not generally preserved by the Heisenberg dynamics. 

Lieb-Robinson bounds address the following simple question: 
Does the Heisenberg dynamics corresponding to, e.g. short range
interactions, satisfy some weaker form of locality?

\subsection{From Locality to Commutators}

Lieb-Robinson bounds are often expressed in terms of commutator estimates.
The relationship between these estimates and the support of local observables
is due mainly to the tensor product structure of the observable algebras. We 
briefly discuss this fact in this section.

Let $\Gamma$ be countable set and $\{ ( \mathcal{H}_x, H_x) \}_{x \in \Gamma}$
denote a collection of quantum systems. Consider two finite sets $X,Y \subset \Gamma$.
If $A \in \mathcal{A}_X$, $B \in \mathcal{A}_Y$, and $X \cap Y = \emptyset$, then
for any finite set $\Lambda \subset \Gamma$ for which $X \cup Y \subset \Lambda$
we can regard $A,B \in \mathcal{A}_{\Lambda}$ and as such $[A,B] = 0$ due to the 
structure of the tensor product. In words, observables with disjoint supports
commute. 

Conversely, Schur's lemma demonstrates the following. If 
$A \in \mathcal{A}_{\Lambda}$ and 
\begin{equation}
[A, \idty_{\Lambda \setminus Y} \otimes B] = 0 \quad \mbox{for all } B \in \mathcal{A}_Y \, ,
\end{equation}
then $\mbox{supp}(A) \subset \Lambda \setminus Y$.
In fact, a more general statement is true. If $A \in \mathcal{A}_{\Lambda}$ almost commutes with all $B \in \mathcal{A}_Y$, then
$A$ is approximately supported in $\Lambda \setminus Y$. The following lemma
appears in \cite{lppl}.

\begin{lemma} \label{lem:app}
Let $\mathcal{H}_1$ and $\mathcal{H}_2$ be Hilbert spaces and $A \in \mathcal{B}( \mathcal{H}_1 \otimes \mathcal{H}_2)$.
Suppose there exists $\epsilon \geq 0$ for which
\begin{equation}
\left\| \left[ A, \idty_1 \otimes B \right] \right\| \leq \epsilon \| B \| \quad \mbox{ for all } B \in \mathcal{B}(\mathcal{H}_2).
\end{equation}
Then, there exists $A' \in \mathcal{B}(\mathcal{H}_1)$, such that
\begin{equation}
\left\| A' \otimes \idty_2 - A \right\| \leq \epsilon.
\end{equation}
\end{lemma}

Equipped with this lemma, we see that uniform estimates on commutators
provide approximate information on the support of observables. Thus, for any
$A \in \mathcal{A}_X$, we can approximate $\mbox{supp}( \tau_t^{\Lambda}(A))$ 
by bounding $\| [\tau_t^{\Lambda}(A),B] \|$ for all $B $ with $\mbox{supp}(B) \subset Y$. Here the
estimates will, of course, depend on the distance between $X$ and $Y$ in $\Gamma$
and the time $t$ for which the observable has been evolved. This is the basic idea of a 
Lieb-Robinson bound.

\section{Lieb-Robinson Bounds}

In this section, we describe in detail assumptions on the set $\Gamma$ and the interactions
$\Phi$ under which one can prove a Lieb-Robinson bound. We also present a precise
statement of the estimate and discuss several relevant consequences. For a proof in this setting, we
refer the interested reader to \cite{Nach09}.

\subsection{On the Geometry of $\Gamma$} \label{sec:geo}

For many models, e.g both Example \ref{ex:HH} and \ref{ex:AH}, the set $\Gamma = \mathbb{Z}^{\nu}$.
In general, though, the lattice structure of $\mathbb{Z}^{\nu}$ is not necessary to prove a Lieb-Robinson bound.
The following assumptions are sufficient. Let  $\Gamma$ be a set equipped with a metric $d$. 
If $\Gamma$ has infinite cardinality, we further assume that there is a non-increasing function 
$F: [0, \infty) \to (0, \infty)$ which satisfies two conditions. First, we assume that $F$ is
{\it uniformly integrable} over $\Gamma$, i.e., 
\begin{equation} \label{eq:uniint}
\| F \| = \sup_{x \in \Gamma} \sum_{y \in \Gamma} F(d(x,y)) < \infty \, .
\end{equation}
Next, we assume there exists $C>0$ such that the following {\it convolution condition} is
satisfied: for all $x,y \in \Gamma$,
\begin{equation} \label{eq:conv}
\sum_{z \in \Gamma} F(d(x,z)) F(d(z,y)) \, \leq \, C \, F(d(x,y)).
\end{equation}
The inequality (\ref{eq:conv}) is quite useful in the iteration scheme which is
at the heart of proving a Lieb-Robinson bound.

Here is an important observation. Let $\Gamma$ be a set with a metric and $F$
satisfy the properties mentioned above with respect to $\Gamma$. In this case, the 
function $F_a$ defined by setting $F_a(r) = e^{-ar}F(r)$ for any $a \geq 0$ 
also satisfies (\ref{eq:uniint}) and (\ref{eq:conv}) above with $\| F_a \| \leq \| F \|$
and $C_a \leq C$. The choice of an exponential weight here is convenient, but
not necessary. In fact, $G = wF$ satisfies (\ref{eq:uniint}) and (\ref{eq:conv})
for any positive, non-increasing, {\it logarithmically super-additive}
weight $w$, i.e. a function $w$ for which $w(x+y) \geq w(x)w(y)$.

\begin{example} Consider the case of $\Gamma = \mathbb{Z}^{\nu}$.
For any $\epsilon >0$, the function $F(r) = (1+r)^{ -( \nu + \epsilon)}$
is positive, non-increasing, and
\begin{equation}
\| F \| = \sum_{x \in \mathbb{Z}^{\nu}} \frac{1}{(1+|x|)^{\nu + \epsilon}} < \infty  .
\end{equation}
A short calculation shows that the convolution constant for this $F$ satisfies 
$C \leq 2^{\nu + \epsilon +1} \| F \|$. Thus, such functions do exist. 
As a final remark, we note that the exponential function {\it does not}
satisfy the convolution condition (\ref{eq:conv}) on $\mathbb{Z}^{\nu}$, however, 
$F_a(r) = e^{-ar}/(1+r)^{\nu+\epsilon}$ certainly does. 
\end{example}

\subsection{Assumptions on the Interaction $\Phi$}

Locality estimates are valid when the interactions are sufficiently short range.
For general sets $\Gamma$, a sufficient decay assumption can be made precise 
in terms of the $F$ function introduced in the previous sub-section. 

Let $\Gamma$ be a set with a metric $d$ and a function $F$ as in Section~\ref{sec:geo}.
For any $a \geq 0$, denote by $\mathcal{B}_a( \Gamma)$ the set of all those interactions 
$\Phi$ for which
\begin{equation}
\| \Phi \|_a \, = \, \sup_{x,y \in \Gamma} \frac{1}{F_a(d(x,y))} \sum_{\stackrel{X \subset \Gamma:}{x,y \in X}} \| \Phi(X) \| \, < \, \infty.
\end{equation}

\begin{example}
Many interesting models have finite range interactions, see e.g. Example~\ref{ex:HH} and \ref{ex:AH}.
An interaction $\Phi$ is said to be of finite range if there exists a number $R>0$ for which 
$\Phi(X) = 0$ whenever the diameter of $X$ exceeds $R$. 
In the case of $\Gamma = \mathbb{Z}^{\nu}$ and $F(r) = (1+r)^{-\nu - \epsilon}$,
it is easy to see that all uniformly bounded, finite range interactions satisfy $\| \Phi \|_a < \infty$ for all $a \geq 0$. 
\end{example}

\begin{example} Another important class of models involve pair interactions. An interaction $\Phi$ is called a pair
interaction if $\Phi(X) = 0$ unless $X= \{ x,y\}$ for some points $x,y \in \Gamma$. 
In the case of $\Gamma = \mathbb{Z}^{\nu}$ and $F(r) = (1+r)^{-\nu - \epsilon}$,
it is easy to see that all uniformly bounded, pair interactions that decay exponentially in $|x-y|$
satisfy $\| \Phi \|_a < \infty$ for some $a > 0$. In fact, if the pair interactions decay faster than an
appropriate inverse polynomial, then $\| \Phi \|_0 < \infty$, and this is sufficient for a decay estimate
on the relevant commutators.
\end{example}

\subsection{The Main Result}

We can now state a Lieb-Robinson bound, proven in \cite{Nach09}, for the systems introduced
above. 

\begin{theorem} \label{thm:lrb}
Let $\Gamma$ be a set with a metric $d$ and a function $F$ as described in Section~\ref{sec:geo}.
Fix a collection of quantum systems  $\{ (\mathcal{H}_x, H_x) \}_{x \in \Gamma}$ over $\Gamma$,
and for any $a \geq 0$, let $\Phi \in \mathcal{B}_a( \Gamma)$. 
Then, the model defined by $\Gamma$, $\{ (\mathcal{H}_x, H_x) \}_{x \in \Gamma}$, and $\Phi$ satisfies
a Lieb-Robinson bound. In fact, for each fixed finite subsets $X,Y \subset \Gamma$,
and any finite $\Lambda \subset \Gamma$ with $X \cup Y \subset \Lambda$, 
the estimate 
\begin{equation} \label{eq:genlrb}
\| [ \tau_t^{\Lambda}(A), B ] \| \leq 2 \| A \| \| B \| \min\left[ 1, g_a(t) \sum_{x \in X, y \in Y} F_a\left( d(x,y) \right) \right]
\end{equation}
holds for any $A \in \mathcal{A}_X$, $B \in \mathcal{A}_Y$, and $t \in \mathbb{R}$.
Here the function $g_a$ is given by
\begin{equation}
g_a(t) = \left\{ \begin{array}{cc} C_a^{-1} \left( e^{2 \| \Phi \|_a C_a |t|} -1 \right) & \mbox{if } X \cap Y = \emptyset, \\
C_a^{-1} e^{2 \| \Phi \|_a C_a |t|} & \mbox{otherwise} . \end{array} \right.
\end{equation}
\end{theorem}
As a corollary, a more familiar form of the Lieb-Robinson bound can be expressed
in terms of $d(X,Y) = \min_{x \in X, y \in Y} d(x,y)$, namely
\begin{corollary} Given the assumptions of Theorem~\ref{thm:lrb} and $a>0$, the estimate
\begin{equation} \label{eq:corlrb}
\| [ \tau_t^{\Lambda}(A), B ] \| \leq  \frac{2 \| A \| \| B \| \| F \|}{C_a} \min[|X|,|Y|] e^{-a \left( d(X,Y) - v_{\Phi}(a) |t| \right)} 
\end{equation}
is valid. Here $|X|$ denotes the cardinality of $X$ and the quantity $v_{\Phi}(a)$ is given by
\begin{equation} \label{eq:vel}
v_{\Phi}(a) = \frac{2 \| \Phi \|_a C_a}{a} \, .
\end{equation}
\end{corollary}

Let us make a few remarks to help interpret these bounds. As we indicated in
Section~\ref{sec:IaM}, the Heisenberg dynamics, $\tau_t^{\Lambda}$, forms a one-parameter
group of automorphisms on $\mathcal{A}_{\Lambda}$ and so the estimate
$\| [ \tau_t^{\Lambda}(A), B ] \| \leq 2 \| A \| \| B\|$ is always true. What we see from
(\ref{eq:genlrb}), and more directly in (\ref{eq:corlrb}), is that if $A$ and $B$ have disjoint supports
$X$ and $Y$ respectively, then $[\tau_0(A), B] = [A,B]=0$ and the
estimate on $[ \tau_t(A), B]$ is small in the distance $d(X,Y)$ for times
\begin{equation}
|t| \leq \frac{d(X,Y)}{v_{\Phi}(a)} \, .
\end{equation}
For this reason, the quantity $v_{\Phi}(a)$ is called a Lieb-Robinson velocity
for the model under consideration. In fact, using Lemma~\ref{lem:app}, we see that
for each $A \in \mathcal{A}_X$ the time evolution $\tau_t^{\Lambda}(A)$ is
approximately supported in a ball of radius $v_{\Phi}(a) |t|$ about $X$.
Thus the dynamics of the system remain essentially confined to a light cone
defined by this Lieb-Robinson velocity. Moreover, the velocity $v_{\Phi}(a)$, see (\ref{eq:vel}),
which governs the rate at which  disturbances propagate through the system, depends
only on the interaction $\Phi$ and the geometry of $\Gamma$; specifically, it is
independent of the on-site Hamiltonians.

Another crucial fact about these Lieb-Robinson bounds is that the explicit estimates, in particular
the velocity, do not depend on the finite volume $\Lambda$ on which the dynamics is defined. 
This suggests, and can be proven in this setting see e.g. \cite{Nach10}, that a thermodynamic limit for the 
dynamics exists. It too satisfies the same Lieb-Robinson bound.

As has been useful in a variety of applications, it is interesting to note the
dependence of these bounds on the support of the corresponding observables.
Since only the minimum cardinality appears, one of two the observables could
be allowed to be volume, i.e. $\Lambda$, dependent without sacrificing the bound. 
In fact, a more detailed analysis shows that only the minimum cardinality of the 
boundary of the supports of the observables is relevant, see \cite{Nach09} for a precise
statement.  

When $\Phi \in \mathcal{B}_a( \Gamma)$ for some $a>0$, the Lieb-Robinson bounds
decay exponentially in the distance between the supports of the observables. The rate of this exponential decay, here the number $a>0$, is usually
of little consequence. For this reason, if $\Phi \in \mathcal{B}_a( \Gamma)$ for  all
$a \in (\alpha, \beta)$, the optimal Lieb-Robinson velocity is given by
\begin{equation}
\inf_{a \in ( \alpha, \beta)} v_{\Phi}(a) = \inf_{a \in ( \alpha, \beta)}  \frac{2 \| \Phi \|_a C_a}{a} \,  .
\end{equation} 
We now estimate this optimum velocity for the previously mentioned examples.

\begin{example} Let $\Gamma = \mathbb{Z}^{\nu}$, $F(r) = (1+r)^{- \nu - \epsilon}$, and
consider the spin 1/2 Heisenberg model introduced in Example~\ref{ex:HH}. Clearly, for any $a >0$,
\begin{equation}
\| \Phi \|_a = e^a 2^{ \nu + \epsilon} 3 J < \infty \, ,
\end{equation}
and therefore, a bound on the optimal velocity of this model is given by
\begin{equation}
3 J e 2^{2( \nu + \epsilon +1)}  \sum_{x \in \mathbb{Z}^{\nu}} \frac{1}{(1+|x|)^{ \nu + \epsilon}} \, .
\end{equation}
As we observed above, this estimate on the velocity is independent of the 
on-site parameter $h$.
\end{example}

\begin{example} Let $\Gamma = \mathbb{Z}^{\nu}$, $F(r) = (1+r)^{- \nu - \epsilon}$, and
consider the anharmonic Hamiltonian introduced in Example~\ref{ex:AH}. A similar calculation
gives a bound on the optimal velocity of
\begin{equation}
\| \Phi \|_{\infty} e 2^{2( \nu + \epsilon +1)}  \sum_{x \in \mathbb{Z}^{\nu}} \frac{1}{(1+|x|)^{ \nu + \epsilon}} \, ,
\end{equation}
which is independent of the on-site function $V$, so long as the self-adjointness
assumption is satisfied.
\end{example}

\section{Some Words on Generalizations and Applications}

Over the past few years applications have driven a number of interesting
generalizations of the original Lieb-Robinson bounds. Several review
articles have been devoted to many of these specific applications, see \cite{NachS09, Eisert08},
and some lecture notes from schools on topics concerning locality are
now available \cite{NachS10, hastings10ln}.
In this short note, we make no attempt to give an exhaustive list of
generalizations and applications, but rather we list many relevant 
works to give the interested reader a reasonable starting point
to further investigate this active area of research.

\subsection{On Generalizations}

The Lieb-Robinson bound stated in Theorem~\ref{thm:lrb}, and proven in \cite{Nach09}, already
includes several generalizations of the original result. Most importantly, it applies
to quantum systems with infinite dimensional, single site Hilbert spaces.
In addition, no assumption on the lattice structure of $\Gamma$ is necessary, and
the dependence of the bound on the support of the observables has been refined.

Recently,  Lieb-Robinson bounds have been proven for time-dependent interactions,
see \cite{lppl}. Moreover, Poulin demonstrated in \cite{Poulin10} that these estimates
also hold for an irreversible, semi-group dynamics generated by Lindblad
operators.   

Quite some time ago, it was proven in \cite{Marchioro78} that the analogue of Lieb-Robinson
bounds hold for the non-relativistic dynamics corresponding to 
classical Hamiltonian systems. In the past few years, further
work in this direction has appeared in \cite{Butta2007} and \cite{Raz09}.

An important open question is: To what extent do Lieb-Robinson bounds
apply in the case of unbounded interactions? For certain simple systems, there
has been some progress on this issue. Lieb-Robinson bounds for
general harmonic systems first appeared in \cite{cramer:2008}. It was proven in 
\cite{Nach09}, see also \cite{Amour09, Nach10}, that these estimates also hold for anharmonic systems if
the perturbation is sufficiently weak. A recent result in \cite{Harrison10}
suggests that such bounds apply much more generally. 
Finally, a Lieb-Robinson estimate for commutator bounded operators appeared in \cite{PS09}. 
 
\subsection{On Applications}
 
Many of the generalizations mentioned above came about by pursuing
concrete applications. As discussed in the main text, the resurgence of
interest in Lieb-Robinson bounds was mainly motivated by 
Hastings' 2004 paper \cite{hastings2004a} on a proof of the multi-dimensional Lieb-Schultz-Mattis 
theorem. In this incredibly influential paper, Hastings discussed generalized
Lieb-Robinson bounds, an Exponential Clustering theorem, and pioneered his
notion of a quasi-adiabatic evolution. This single work inspired a flurry of
activity which continues to this day.

The Lieb-Schultz-Mattis theorem, see \cite{lieb1961}, concerns the
spectral gap between the ground state energy and that of the
first excited state for the nearest-neighbor, spin 1/2 Heisenberg 
model in one dimension. They proved that for a finite volume of
size $L$, if the ground state is unique, then the gap is bounded
by $C/L$, for some constant $C$. Further generalizations, to
models with arbitrary half-integer spin and to a statement valid
in the thermodynamic limit appeared in \cite{AL86}. Hastings paper \cite{hastings2004a}
developed a multi-dimensional analogue of this result. In fact, his
argument yields a gap estimate applicable in a great deal of generality, 
see \cite{Nach07} for a precise statement. Recent reviews of these results appear in
\cite{NachS09} and \cite{hastings10ln}. 

The Exponential Clustering theorem is a proof that the ground state
expectations of gapped systems decay exponentially in space. 
Proofs of this result first appeared in \cite{NachS06} and \cite{hastings06}. A refinement
of the dependence of the estimates on the support of the observables
was proven in \cite{NachS09}, and this fact was later used by Matsui in \cite{matsui:2010}
to investigate a split property for quantum spin chains.

It is well known that, for quantum spin systems,  a Lieb-Robinson bound
may be used to establish the existence of a thermodynamic limit for
the Heisenberg dynamics, see e.g. \cite{bratteli1997}. Improved estimates allowed for
this result to be generalized, e.g., the case of polynomially decaying
interactions was covered in \cite{Nach06} and the existence of the dynamics
for the general systems considered here was proven in \cite{Nach10}.
For perturbations of the harmonic system, the existence of the thermodynamic
limit has been proven with two distinct methods, see \cite{Amour09} and \cite{Nach10}. 

An area law for gapped one-dimensional systems was proven by Hastings in \cite{hastings2007}.
In general, the area law conjecture states that the von Neumann entropy of the
restriction of gapped ground states to a finite volume of size $\Lambda$
grows no faster than a quantity proportional to the surface area of $\Lambda$. 
 Certain aspects of Hastings' argument generalize to the multi-dimensional
 setting, e.g. a factorization property of gapped ground states was proven in
 \cite{Hamza09}, but a proof of the area law for general gapped systems in arbitrary
 dimension remains an important open question. Some progress on a 
 class of unfrustrated spin Hamiltonians appears in \cite{debeaudrap10}. A review of these
 topics is contained in \cite{Eisert08}, see also \cite{NachS10}.  
 
Quantization of the Hall conductance for a general class of interacting 
fermions was recently proven in \cite{hastings09}. This intriguing result makes crucial
use of improved Lieb-Robinson bounds and the methods associated
with Hastings' quasi-adiabtic evolution. A detailed analysis of this technique,
with specific regards to its implications for perturbation theory, is the main topic of \cite{lppl}. 

Finally, stability of topological order was addressed in \cite{bravyi:2010a, bravyi:2010b}. There the authors
consider a class of Hamiltonians that are the
sum of commuting short-range terms, such as the toric code model developed by
Kitaev in \cite{kitaev2003}, and proved that the topological order of the ground states
is stable under arbitrary, small short-range perturbations.

Developing a better understanding of quantum dynamics and its
perturbation theory will be crucial in providing new insight into 
complex physical phenomena.  As indicated by the number of 
recent generalizations and applications, the
analysis of Lieb-Robinson bounds is a thriving area of active research
which attempts to address this very issue. 

\section*{Acknowledgments}
I would like to thank my collaborator Bruno Nachtergaele for many
inspiring and insightful discussions which have greatly enriched my
understanding of this field and strongly influenced the better parts of this presentation. 
The work reported in this paper was supported by the National Science Foundation
under grants \#DMS-0757581 and \#DMS-0757424.

\end{document}